\documentclass[aps,pra,floatfix,superscriptaddress,showpacs,twocolumn]{revtex4}
\usepackage{amsmath}
\usepackage{graphicx}
\def\prn#1{{\left(#1\right)}}

\newcommand{\abs}[1]{\left|#1\right|}

\def\bra#1{{\left\langle#1\right\vert}}
\def\ket#1{{\left\vert#1\right\rangle}}

\def\prn#1{{\left(#1\right)}}
\def\brk#1{{\left[#1\right]}}

\def\abs#1{{\left|#1\right|}}

\newcommand{\sixS}{${\rm 6S}_{1/2}\,$}
\newcommand{\eightS}{${\rm 8S}_{1/2}\,$}
\newcommand{\nineS}{${\rm 9S}_{1/2}\,$}
\newcommand{\sevenDthree}{${\rm 7D}_{3/2}\,$}
\newcommand{\sevenDfive}{${\rm 7D}_{5/2}\,$}
\newcommand{\sixPone}{${\rm 6P}_{1/2}\,$}
\newcommand{\sixPthree}{${\rm 6P}_{3/2}\,$}

\def\threeJ#1#2#3#4#5#6{\biggl( \begin{array}{ccc} #1 & #2 & #3 \\ #4 & #5 & #6 \end{array} \biggr)}
\def\sixJ#1#2#3#4#5#6{\biggl\{ \begin{array}{ccc} #1 & #2 & #3 \\ #4 & #5 & #6 \end{array} \biggr\}}

\newcommand{\nist}{Time and Frequency Division, National Institute of Standards and
Technology, 325 Broadway MS 847, Boulder CO 80305}
\newcommand{\witwatersrand}{School of Physics, University of Witwatersrand, Private Bag 3, WITS, 2050, RSA}
\newcommand{\oberlin}{Department of Physics and Astronomy, Oberlin College, 110 N. Professor St., Oberlin, OH 44074}
\newcommand{\notredame}{Department of Physics, University of Notre Dame, Notre Dame, Indiana 46556-5670}

\begin{document}
\title{Femtosecond frequency comb measurement of absolute frequencies and hyperfine coupling constants in cesium vapor}
\author{Jason E.\ Stalnaker\footnote{Corresponding author:
jason.stalnaker@oberlin.edu}} \affiliation{\oberlin}
\affiliation{\nist}
\author{Vela Mbele\footnote{Present address: P. O. Box 755, Durban, KWAZULU-NATAL, 4000, RSA.}} \affiliation{\nist}
\affiliation{\witwatersrand}
\author{Vladislav Gerginov\footnote{Present address: Physikalisch-Technische Bundesanstalt, Bundesallee 100, 38116, Germany}}
\affiliation{\nist}
\author{Tara M.\ Fortier}
\affiliation{\nist}
\author{Scott A.\ Diddams}
\affiliation{\nist}
\author{Leo Hollberg \footnote{Present address: P.O.\ Box 60157, Sunnyvale, CA 94088}}
\affiliation{\nist}
\author{Carol E.\ Tanner}
\affiliation{\notredame}

\date{\today}
\begin{abstract}
We report measurements of absolute transition frequencies and
hyperfine coupling constants for the \eightS, \nineS, \sevenDthree,
and \sevenDfive states in $^{133}{\rm Cs}$ vapor.  The stepwise
excitation through either the \sixPone or \sixPthree intermediate
state is performed directly with broadband laser light from a
stabilized femtosecond laser optical-frequency comb.  The laser beam
is split, counter-propagated and focused into a room-temperature Cs
vapor cell. The repetition rate of the frequency comb is scanned and
we detect the fluorescence on the $7{\rm P}_{1/2,3/2} \rightarrow
6{\rm S}_{1/2}$ branches of the decay of the excited states.  The
excitations to the different states are isolated by the introduction
of narrow-bandwidth interference filters in the laser beam paths.
Using a nonlinear least-squares method we find measurements of
transition frequencies and hyperfine coupling constants that are in
agreement with other recent measurements for the 8S state and
provide improvement by two orders of magnitude over previously
published results for the 9S and 7D states.

\end{abstract} \pacs{42.62.Fi, 42.62.Eh, 32.30.-r, 32.10.Fn}


\maketitle
\section{Introduction}

The basic ideas of using repetitive pulses and mode-locked lasers
for high-resolution precision spectroscopy were originally developed
in the 1970's, with particular focus on Doppler-free two-photon
transitions
\cite{teets77,baklanov77,eckstein78,salour78,hansch75,hansch80}. The
advent of fully stabilized optical frequency combs \cite{jones00,
holzwarth00} in 2000 renewed interest in many spectroscopic
applications of mode-locked lasers (for a review see Ref.\
\cite{stowe08}).  In many instances the use of direct frequency-comb
spectroscopy (DFCS) offers many advantages over spectroscopy done
with cw lasers.  In particular, the spectral versatility provided by
modern frequency combs allows the simultaneous study of multiple
atomic transitions with a single experimental setup.
Additionally, DFCS allows for a relatively simple determination of
the absolute frequency by providing a direct link to the S.I.
second. In this work we exploit this versatility to study multiple
transitions in cesium (Cs).

Precision spectroscopy of Cs plays an important role in atomic and
fundamental physics, including measurements of atomic parity
violation \cite{wood97,bennett99, guena05}, searches for the
permanent electric dipole moment of an electron
\cite{murthy89,nataraj08}, photon recoil measurements that help
determine the fine structure constant
\cite{wicht02,gerginov06}, measurements of the
distribution of nuclear charge and magnetization \cite{gerginov03},
and numerous applications in laser cooling, Bose-Einstein
condensation \cite{leggett01}, atomic clocks \cite{wynards05}, and
magnetometers \cite{budker07}.

The present work focuses on measuring multiple transition
frequencies and the hyperfine structure of select low-lying states in cesium via two-photon excitation.  The relatively simple experimental setup uses a
room-temperature Cs vapor cell and a self-referenced optical
frequency comb to excite Doppler-free two photon transitions from
the Cs ground state to the \eightS, \nineS, \sevenDthree, and
\sevenDfive states.  Our results are consistent with previous
measurements for the well-studied \eightS state
\cite{hagel99,fendel07} and provide significantly reduced
uncertainties for the \nineS, \sevenDthree, and \sevenDfive states
\cite{weber87}.

This paper is structured as follows: Section \ref{sec expt} contains a description of our experimental system and measurements.  In Section \ref{sec theory} we examine the features of the fluorescence spectra and discuss the theoretical model used to describe the results.  Section \ref{sec analysisresults} describes the analysis  of the two-photon transitions and the recipe used in extracting the transition energies.  This section also includes our results and the details of the systematic effects.  We conclude our
work in Section \ref{sec conclusions}.

\section{Experimental Setup}\label{sec expt}

The relevant Cs energy levels are shown in Fig.\ \ref{fig
csenergydiagram}. The excitation pathways to the four states
\eightS, \nineS, \sevenDthree, or \sevenDfive proceed through a
stepwise excitation through either the \sixPone or \sixPthree
states.  We detect the two-photon excitation as blue fluorescence
emission as the atoms decay from the $7{\rm P}_{1/2,3/2}$ states to
the ground state.

\begin{figure}
\centerline{\includegraphics[width=3.5 in]{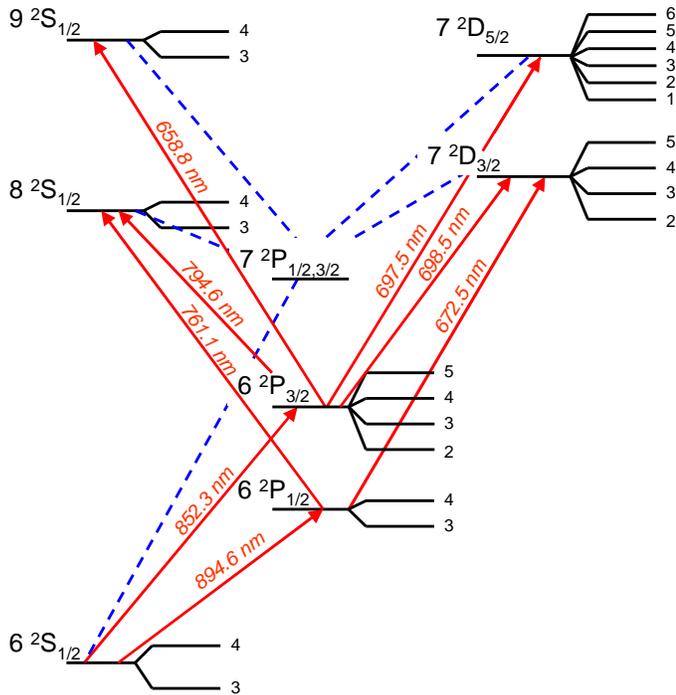}}
\caption{(Color online) Energy levels of
$^{133}\textrm{Cs}\: (I = 7/2)$ that are relevant to this work. The excitation pathways are shown in
solid red along with the transition wavelengths. The decay channels
used in the detection are shown in blue dashed lines. The total
angular momentum of the hyperfine components is listed next to each
state. } \label{fig csenergydiagram}
\end{figure}

\begin{figure}
\centerline{\includegraphics[width=3.5 in]{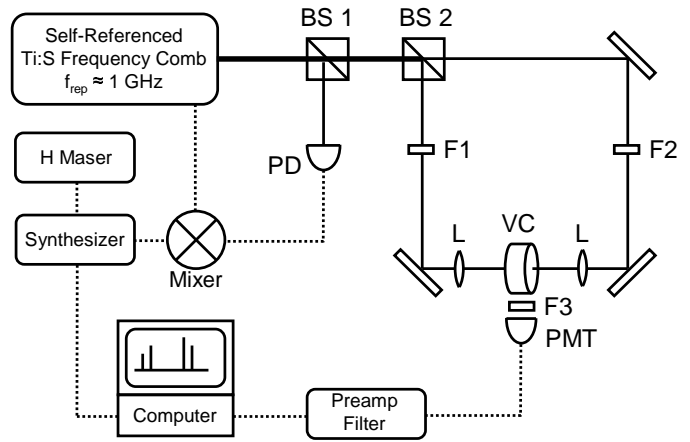}}
\caption{Block diagram of the experimental set up.  The components
are: BS1, BS2 nonpolarizing beam splitters; F1, F2, F3 interference
filters; L lens; PD photodiode: PMT photomultiplier tube; VC: Cs
vapor cell.} \label{fig exptsetup}
\end{figure}

The experimental set-up is shown in Fig.\ \ref{fig exptsetup}.  We
excite the cascaded two-photon transitions using a mode-locked
Ti:Sapphire laser that generates a broad optical frequency comb. The
portion of the output spectrum that was used extended from 650 nm to
1000 nm, covering all of the excitation transitions of interest in
Fig.\ \ref{fig csenergydiagram}. The laser is described in detail in
Ref.\ \cite{fortier06a} and here we only briefly mention the
information relevant for the experiment. The Kerr-lens mode-locked
Ti:Sapphire ring laser produces an output spectrum consisting of
discrete equally spaced coherent optical modes.  The frequency of
each optical mode, $\nu_n$, can be related to two microwave
frequencies, the laser repetition rate, $f_{rep}$, and the
carrier-envelope offset frequency, $f_0$, by
\begin{align}
\nu_n = n\, f_{rep}+f_0,
\end{align}
where $n$ is an integer mode number $\sim 4 \times 10^5$.  The
carrier-envelope-offset frequency is stabilized by use of an
$f$-to-$2f$ interferometer, thus ``self-referencing" all modes of
the comb \cite{jones00}. For the measurements described here $f_0$
was stabilized to 45 MHz. The repetition rate of the laser is
$\approx 1$ GHz.  The repetition rate was phase-locked to a signal
generator that was referenced to a hydrogen maser.  With the comb
stabilized in this way, the fractional instability of every optical
frequency of the comb is $\approx 2\times 10^{-13}\,
\tau^{-\frac{1}{2}}$, where $\tau$ is the averaging period in
seconds.

The laser beam was split by use of a 50-50 nonpolarizing
beam-splitting cube (BS 2).  The two output laser beams were
linearly polarized along the same axis. The beams were filtered by use of either narrow-band interference filters or a combination of
long-pass and short-pass filters to select specific transition
pathways. The counter-propagating beams were then focused by lenses of 25 cm focal length to a spot size of $\sim 100$ $\mu$m inside the vapor cell. Focusing the beams to a smaller waist size increases the two-photon excitation rate, but also leads to broadening of the resonances due to the finite transit-time of the atoms through the beams.  The transit-time broadening for a waist of this size is about 2 MHz for room-temperature Cs atoms. The average optical intensities in the vapor cell were tens of ${\rm W/cm^2}$ spread over $\approx 10$ nm bandwidth, corresponding to $\approx 4000$ optical frequencies.


An evacuated Pyrex Cs vapor cell was used in the experiment.  The
cell has a diameter of 2.5 cm and is a wedged shape with an optical
path length ranging from 0.5 cm - 1 cm.  Windows were fused directly
to the cell at an angle.  The cell was carefully cleaned and filled
on an oil-free vacuum system and was evacuated to a base pressure of
$\sim 10^{-5}$ Pa.  A small amount of Cs was introduced into the
cell prior to removing it from the vacuum system.  The cell
temperature was varied by resistive heater wires wound around the
cell.  We note that we observed shifts in the transition frequencies
as large as 0.5 MHz in a Cs vapor cell that was not cleaned and
vacuum processed as carefully as the cell used in these
measurements.

Most of the measurements in this experiment were taken with the Cs
cell temperature at 297 K, with fluctuations less than 1
K.  Data were taken at higher temperatures to evaluate
systematic shifts arising from pressure shifts as described in
Section \ref{sec 8Ssyst}.

A set of six coils was used to zero out the earth's magnetic field
in the region of the vapor cell.  The field was reduced to a level
of $<1 \: \mu\textrm{T}$ over the interaction region inside the
vapor cell.

Two-photon resonance signals were monitored by measuring the
$7\textrm{P}_J \rightarrow 6\textrm{S}_{1/2}$ ($J=1/2$, $3/2$)
cascade fluorescence at 459 nm and 455 nm.  The fluorescence was
collected at a right angle to the two counter-propagating beams and
was detected by a photomultiplier tube.  The detector was shielded
from background light and the scattered light from the excitation
laser beams by a 457.9 nm filter with a 10 nm pass band (F3).  The
detector output was amplified and filtered (low-pass at 1 kHz, 6
dB/octave) before being sampled and recorded with a
computer-controlled data acquisition board (DAQ).

The DAQ used in these measurements is similar to that used in Refs.\
\cite{gerginov06,gerginov03,gerginov04}.  The computer controlled
the scan of the repetition rate by stepping the frequency of the
signal generator to which the repetition rate was phase locked. Data
scans were collected by changing the frequency of the signal
generator from $1000.771\, 725$ MHz to $1000.774 \, 725$ MHz in 1 Hz
steps.  This 3 kHz range in the repetition rate results in a 2-3 GHz
variation in the two-photon frequency, depending on the transition.
At each 1 Hz step the fluorescence was digitized with multiple
samples and averaged for 1 s.  The statistical mean and the
uncertainty in the mean were evaluated at each repetition rate
frequency.  Spectra were collected for both increasing and
decreasing frequency scans, resulting in 6002 data points.

\section{Observed Spectra and Theoretical Considerations}\label{sec theory}

\begin{figure}
\centerline{\includegraphics[width=3.5 in]{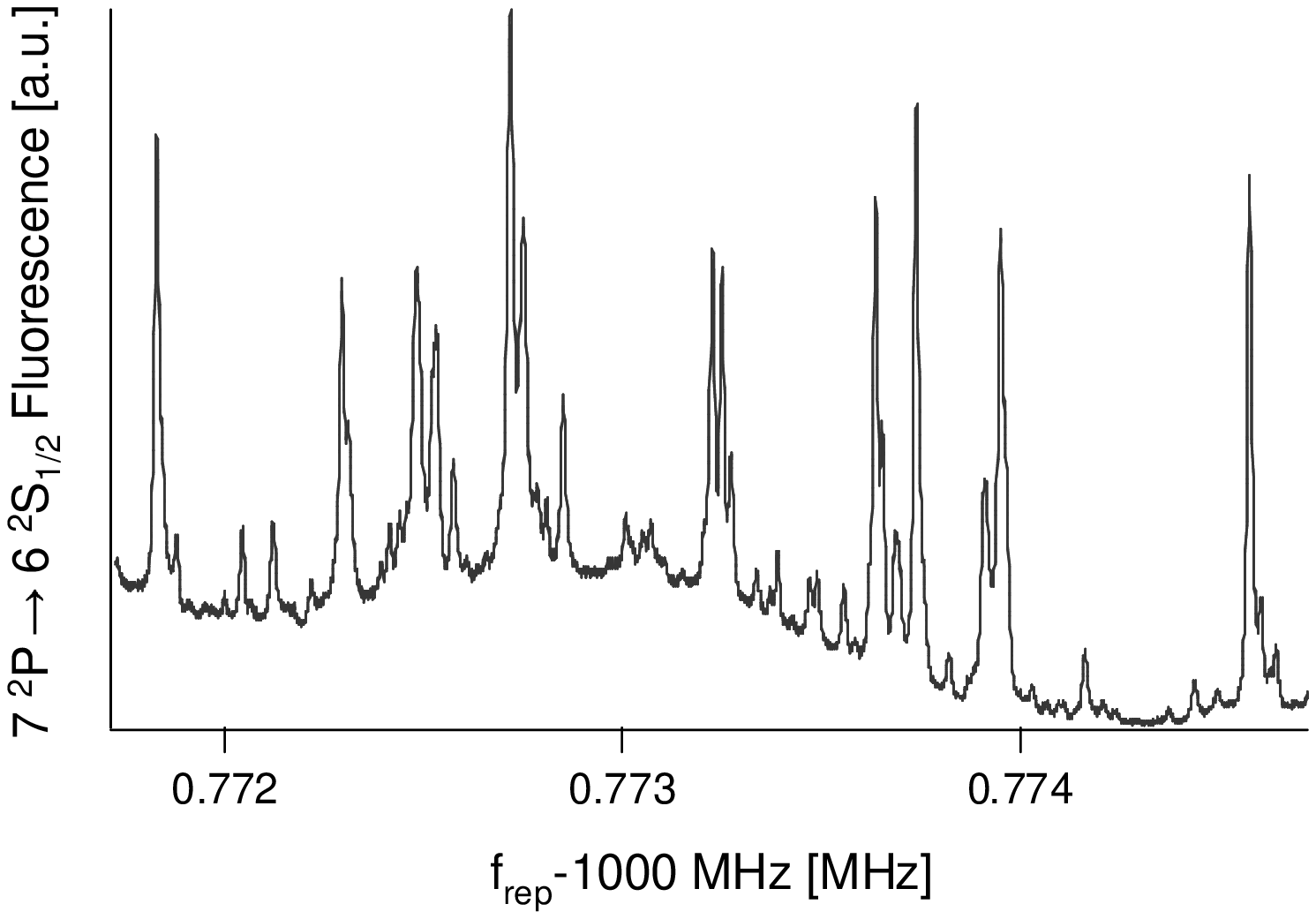}}
\caption{Fluorescence from the
$\textrm{7P}_J\rightarrow\rm{6S}_{1/2}$ decay as a function of laser
repetition rate when no filters (F1 and F2 in Fig.\ \ref{fig
exptsetup}) are used to restrict the laser's spectral bandwidth.}
\label{fig fullSpec}
\end{figure}

Figure \ref{fig fullSpec} shows the spectrum collected as the
repetition rate of the frequency comb was scanned without using the
filters marked as F1 and F2 in Fig.\ \ref{fig exptsetup}.  The
spectrum consists of a multitude of narrow Doppler-free peaks and a
Doppler-broadened background.  The narrow peaks are due to
excitation to the 8S, 9S and 7D states by two counter-propagating
photons, while the Doppler-broadened background is due to excitation
from two co-propagating photons.  The observed spectrum is quite
complicated because of the numerous discrete frequencies present in
the comb combined with the multiple atomic energy levels and
transition pathways accessible.

\begin{figure}
\centerline{\includegraphics[width=3.5 in]{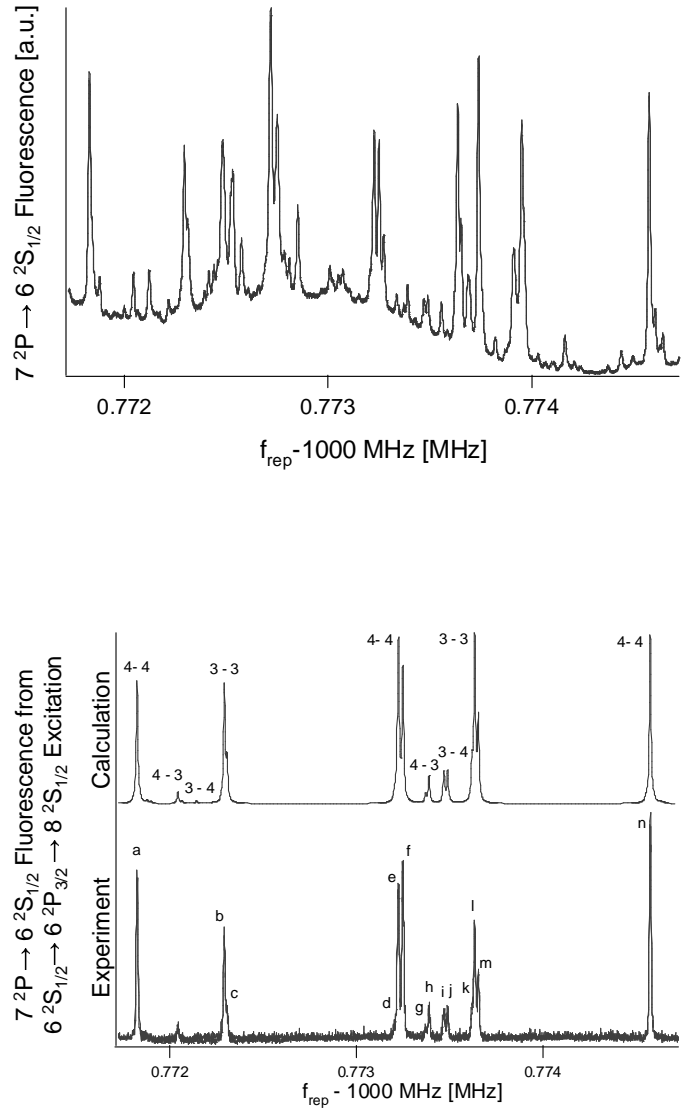}}
\caption{The calculated spectrum (upper trace) and experimental
signal (lower trace) of the $7\textrm{P} \rightarrow 6\textrm{S}$
fluorescence when the Cs atoms are excited to the \eightS state
through the \sixPthree state.  The groups of peaks in the calculated
spectrum are labeled by $F$-$F^{\prime\prime}$ to denote the
hyperfine ground state and hyperfine final state of the transition.  The peaks in each
group correspond to the different hyperfine components of the
intermediate state.  The peaks in the experimental spectrum that are
labeled with letters were used in extracting the transition
frequencies, as described in the text.} \label{fig 8SD2Spec}
\end{figure}

\begin{figure}
\centerline{\includegraphics[width=3.25 in]{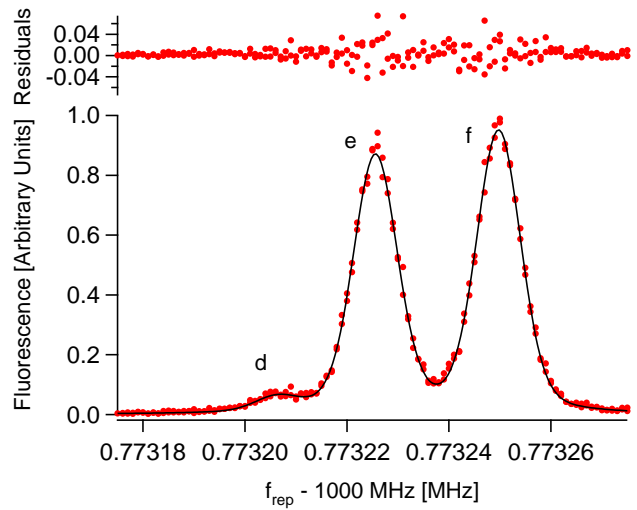}}
\caption{The $7\textrm{P} \rightarrow 6\textrm{S}$ fluorescence when
the Cs atoms are excited to the \eightS state through the \sixPthree
state.  The peaks correspond to the $d$, $e$, and $f$ peaks in Fig.\
\ref{fig 8SD2Spec}.  This data set was taken at a temperature of
319 K and was used in evaluating systematic effects related
to temperature dependence.  } \label{fig 8SD2Sample}
\end{figure}

The spectrum is greatly simplified and the Doppler background is
eliminated by introducing filters F1 and F2, as shown in Fig.
\ref{fig exptsetup}.  The presence of the filters restricts the
transitions and enables the selection of individual excitation
pathways.  Figure \ref{fig 8SD2Spec} shows the spectrum observed
when one of the counter-propagating laser beams is restricted to be
between 845 nm to 855 nm and the other between 780nm to 800 nm. With this combination, the first beam can excite the atoms on only the
$6{\rm S}_{1/2}\rightarrow 6{\rm P}_{3/2}$ transition at 852 nm (the $D_2$ transition) and the second beam can only excite the atoms on
the $6{\rm P}_{3/2}\rightarrow 8{\rm S}_{1/2}$ transition at 795 nm. This eliminates excitations to the \nineS, \sevenDthree, and
\sevenDfive states, as well as the excitation to the \eightS state
through the \sixPone state and removes the Doppler-broadened
background.  The resulting spectrum consists of groups of two or
three peaks.  The peaks labeled $d$, $e$, and $f$ form one such
spectral group and are shown in more detail in Fig.\ \ref{fig
8SD2Sample}. Each spectral group is the result of excitation from a
given ground-state hyperfine level to a given excited-state
hyperfine level, and the individual peaks in the group correspond to specific transition pathways through every allowable hyperfine level of the intermediate state.  For example, the $d$, $e$, and $f$ peaks in Fig.\ \ref{fig 8SD2Sample} all correspond to excitation from the
the \sixS($F= 4$) ground state to the \eightS($F^{\prime\prime}= 4$) excited state.  The three peaks correspond to excitation through the \sixPthree($F^\prime = 3, \, 4,\, 5$) intermediate states.  We see
the groups recur as a different pair of optical modes is resonant
with a given transition. This repeating pattern occurs at intervals
of $\delta f_{rep} \approx 1.2$ kHz, corresponding to a change in
the two-photon optical frequency of $\approx 1$ GHz (i.e., the laser mode spacing).  For a given group of three peaks there are three
different velocity classes that satisfy the resonance condition for
each of the corresponding hyperfine components of the intermediate
level. For another group of three peaks, the resonant condition for
the components of the intermediate level is satisfied by a different pair of comb modes, and by different velocity classes.  Thus, the
amplitudes of the peaks arising from the same ground, excited, and
intermediate states are not the same in the different spectral
groups.

%

We can understand and calculate the spectra using the standard
two-photon transition probabilities derivable from second-order
time-dependent perturbation theory. We start by considering a
transition of an atom in the ground state with total angular
momentum $F$, $\ket{6\textrm{S}_{1/2} \: F}$,  to an excited state
with total angular momentum $F^{\prime\prime}$,
$\ket{n^{\prime\prime}\, L^{\prime\prime}_{J^{\prime\prime}}\:
F^{\prime\prime}}$ via excitation by two cw lasers with angular
frequencies $\omega_1$ and $\omega_2$, intensities $I_1$ and $I_2$,
and wave vectors $\vec{k}_1$ and $\vec{k}_2$.  Here
$n^{\prime\prime}$ refers to the principal quantum number of the
excited state.  We further assume that $\omega_1$ is close to the
resonant frequency for the $6\textrm{S}_{1/2}\rightarrow
6\textrm{P}_{J^\prime}$ transition, with $J^\prime = 1/2$ or $3/2$.
The probability of the atom to make a transition is
\cite{demtroder03}
\begin{widetext}
\begin{small}
\begin{align}
P\prn{6\textrm{S}_{1/2} F ,\, n^{\prime\prime}
L^{\prime\prime}_{J^{\prime\prime}} F^{\prime\prime}} \propto &
\frac{1}{2F+1} \frac{I_1\, I_2}{\brk{\omega_{6\textrm{S}_{1/2} F:
n^{\prime\prime}\, L^{\prime\prime}_{J^{\prime\prime}}
F^{\prime\prime}}-\prn{\omega_1+\vec{k_1}\cdot\vec{v}}-\prn{\omega_2+\vec{k_2}\cdot\vec{v}}}^2+\prn{\frac{\gamma_{n^{\prime\prime}\,
L^{\prime\prime}_{J^{\prime\prime}}}}{2}}^2} \nonumber \\
& \times \sum_{M_F,\, M_F^{\prime\prime}}
\abs{\sum_{F^\prime,M_F^\prime}\frac{\bra{n^{\prime\prime}L^{\prime\prime}_{J^{\prime\prime}}F^{\prime\prime}
M_F^{\prime\prime}}\hat{e}_2\cdot\vec{d}\ket{6\textrm{P}_{J^\prime}F^\prime
M_F^\prime}\bra{6\textrm{P}_{J^\prime}F^\prime
M_F^\prime}\hat{e}_1\cdot\vec{d}\ket{6\textrm{S}_{1/2}F
M_F}}{\omega_{6\textrm{S}_{1/2} F: 6\, \textrm{P}_{J^{\prime}}
F^{\prime}}-\prn{\omega_1+\vec{k}_1\cdot\vec{v}}-i\,\frac{\gamma_{6\textrm{P}_{J^\prime}}}{2}}}^2
, \label{eq twophotonprob}
\end{align}
\end{small}
\end{widetext}
where $\vec{v}$ is the velocity of the atom, $M_F$, $M_F^\prime$,
and $M_F^{\prime\prime}$ are the projections of the total angular
momenta $F$, $F^\prime$, and $F^{\prime\prime}$ along the axis of
quantization, $\vec{d}$ is the electric dipole operator,
$\hat{e}_{1(2)}$ is the unit vector along the direction of the
polarization for the first(second) laser beam, $\gamma_{n \, L_J}$
is the homogeneous line width of the state $\ket{n \, L_J}$, and
$\omega_{n\, L_{J} F: n^{\prime}\, L^{\prime}_{J^{\prime}}F^\prime}$
is the resonant angular frequency of the transition $\ket{n\, L_{J}
\, F} \rightarrow \ket{ n^{\prime}\, L^{\prime}_{J^{\prime}}\,
F^\prime}$. For a given light polarization, specified by the vector
component in the spherical basis $q$, the matrix elements for the
different magnetic sublevels can be related to reduced matrix
elements that do not depend on the magnetic sublevels or the total
angular momentum ($F$, $F^\prime$, or $F^{\prime\prime}$) by use of
the Wigner-Eckart theorem and standard Clebsch-Gordan relations as
\cite{sobelman96}
\begin{widetext}
\begin{align}
\bra{n^{\prime}L^{\prime}_{J^{\prime}}F^{\prime} M^{\prime}_F}d_q
\ket{n L_J F M_{F}} & =
(-1)^{F^\prime-M_F^\prime}\threeJ{F^\prime}{1}{F}{-M_F^\prime}{q}{M_F}
\, \prn{n^{\prime}L^{\prime}_{J^{\prime}}F^{\prime} ||d ||n L_{J}F }
\nonumber \\ & =
(-1)^{F^\prime-M_F^\prime+I+1+J+F}\sqrt{\prn{2F+1}\prn{2F^\prime+1}}\threeJ{F^\prime}{1}{F}{-M_F^\prime}{q}{M_F}
\,\nonumber \\ & \hspace{2in} \times \sixJ{J^\prime}{F^\prime}{I}{F}{J}{1} \prn{n^\prime
L^\prime_{J^\prime} ||d ||n L_J }.
\end{align}
\end{widetext}
Here $I=\frac{7}{2}$ is the nuclear spin and the
$\bigl(\;\;\;\;\;\;\bigr)$ and $\bigl\{\;\;\;\;\;\;\bigr\}$ terms
are the usual 3-$J$ and 6-$J$ symbols.  Thus, the transition
probabilities for different ground-state angular momenta $F$ and
excited-state angular momenta $F^{\prime\prime}$ can be related to a
common factor by use of the appropriate Clebsch-Gordan coefficients.
If we assume the two laser beams are counter-propagating then
examination of Eq.\ \ref{eq twophotonprob} reveals that for a given
transition from the ground state with angular momentum $F$, through
an intermediate state with angular momentum $F^\prime$, to a final
state with $F^{\prime\prime}$ the transition probability will be a
maximum when
\begin{align}
\omega_1\prn{1+\frac{v}{c}}+\omega_2\prn{1-\frac{v}{c}} &
=\omega_{6\textrm{S}_{1/2} F: n^{\prime\prime}\,
L^{\prime\prime}_{J^{\prime\prime}} F^{\prime\prime}}
\label{eq twophotonres}\\ &\textrm{and} \nonumber \\
\omega_1\prn{1+\frac{v}{c}} &  =  \omega_{6\textrm{S}_{1/2} F: 6\,
\textrm{P}_{J^{\prime}} F^{\prime}} .\label{eq onephotonres}
\end{align}
If we now consider the effect of having a number of discrete
frequencies, coming from the different modes of the comb, we see
that, as a result of the detuning from the intermediate state, only
two near-resonant comb modes will contribute substantially to the
transition probability.  Because the spacing of the frequencies for
the comb, $f_{rep} \approx 1$ GHz, is much larger than the
homogeneous linewidths of the states, $\frac{\gamma}{2 \pi}\approx
5$ MHz, we are justified in considering only the two modes closest
to the resonance conditions of Eqs.\ \eqref{eq twophotonres} and
\eqref{eq onephotonres}.  Since these two frequencies are coming
from the comb, they are related to the offset frequency and
repetition rate of the comb by two integers as
\begin{align}
\omega_1 = &\, 2\pi\prn{f_0+n_1\, f_{rep}} \label{eq laserFreq1}\\
\omega_2 = & \, 2\pi\prn{f_0+n_2\, f_{rep}}  \label{eq laserFreq2}.
\end{align}
Substituting these relations into Eqs.\ \eqref{eq twophotonres} and
\eqref{eq onephotonres} we see that for a given pair of modes, $n_1$
and $n_2$, and a fixed value of the offset frequency there exists a
unique repetition-rate frequency and velocity that will
simultaneously satisfy both resonance conditions.  Therefore, given
an atomic sample with a broad enough velocity distribution, there
will exist a repetition-rate frequency that will result in a
stepwise resonant excitation of the two-photon transition for a
specific velocity class.  This leads to the narrow Doppler-free
peaks seen in Fig.\ \ref{fig 8SD2Spec}. Because the velocity class
excited and the resonant repetition-rate frequency both depend on
the energy of the intermediate state, the transition resonance for
each different intermediate state occurs at different repetition
rates. Because a different velocity class contributes to the
different peaks, the amplitudes of the peaks vary greatly, depending
on the number of atoms that are present in the resonant velocity
class.


\section{Analysis and Results}\label{sec analysisresults}
\subsection{Analysis}\label{sec analysis}

We calculated the excitation probabilities by numerically
integrating Eq.\ \ref{eq twophotonprob} over a Gaussian velocity
distribution for optical modes near the resonant condition.  A total of $\approx 15$ optical modes were used (10 for the first stage
excitation and 5 for the second stage excitation). The calculated
spectra agree quite well with the experimental spectra (Fig.\
\ref{fig 8SD2Spec}). Slight variations in the relative amplitudes of the peaks are attributed to slight deviation of the polarization of
the excitation light from linear.

The resonant frequencies for a transition from a ground state with a specific $F$ to an excited state with $F^{\prime\prime}$ are related to the center-of-gravity frequencies and the hyperfine coupling
constants through (see e.g. \cite{corney79})
\begin{widetext}
\begin{align}
\nu_{6{\rm S}_{1/2}F: n^{\prime\prime} L^{\prime\prime}_{J^{\prime\prime}} F^{\prime\prime}} &=\nu_{6{\rm
S}_{1/2}: n^{\prime\prime}L^{\prime\prime}_{J^{\prime\prime}}} - \frac{1}{2}A_{6S_{1/2}}\, \kappa\prn{F,J} +
\frac{1}{2}A_{n^{\prime\prime}L^{\prime\prime}_{J^{\prime\prime}}}\,
\kappa\prn{F^{\prime\prime},J^{\prime\prime}} \nonumber \\ &
+B_{n^{\prime\prime}L^{\prime\prime}_{J^{\prime\prime}}}
\frac{\frac{3}{4}\kappa\prn{F^{\prime\prime},J^{\prime\prime}}^2
+\frac{3}{4}\kappa\prn{F^{\prime\prime},J^{\prime\prime}}-
I\prn{I+1}J^{\prime\prime}\prn{J^{\prime\prime}+1}}
{2I\prn{2I-1}J^{\prime\prime}\prn{2J^{\prime\prime}-1}}, \label{eq
energies}
\end{align}
\end{widetext}
where $\kappa\prn{F,J}=F\prn{F+1}-I\prn{I+1}-J\prn{J+1}$.

The repetition rates at which the calculated resonances occur depend on the center-of-gravity frequencies for the two stages of the
excitation and the hyperfine $A$ and $B$ constants of the ground,
intermediate, and final states.  Spectra were calculated with
different values of the two-photon center-of-gravity frequency and
hyperfine coupling constants of the final excited state. The
intermediate state energy and coupling constants used in the
calculated spectra were fixed to the values determined from Gerginov et al. \cite{gerginov04,gerginov06}.  The two-photon transition
center-of-gravity frequencies $\nu_{6{\rm S}_{1/2}: n^{\prime\prime} L^{\prime\prime}_{J^{\prime\prime}}}$ and the hyperfine coupling
constants $A_{n^{\prime\prime} L^{\prime\prime}_{J^{\prime\prime}}}$
and $B_{n^{\prime\prime} L^{\prime\prime}_{J^{\prime\prime}}}$ were
extracted from the experimental data by least-squares comparison of
the experimental resonant repetition-rate frequencies with the
resonant repetition-rate frequencies determined from the calculated
spectra.

The resonant repetition rate for each of the peaks in both the
experimental and calculated spectra were determined from a
least-squares fit.  Fitting the spectra was necessary in order to
account for the overlap of the different peaks.  The calculated
spectra were assumed to have a Lorentzian line shape. However, the
experimental spectra were found to be well described by a line shape of a Gaussian coupled to a Lorentzian,
\begin{align}
\sum_i\prn{C^G_i e^{-\frac{\prn{f-f_i}^2}{2\,
\gamma_G^2}}+C^L_i\frac{\prn{\frac{\gamma_L}{2}}^2}
{\prn{f-f_i}^2+\prn{\frac{\gamma_L}{2}}^2}} +C^B ,
\end{align}
where $C^G_i$ and $C^L_i$ are the Gaussian and Lorentzian
amplitudes, respectively, $\gamma_G$ and $\gamma_L$ are the Gaussian and Lorentzian widths, $f_i$ is the line center of the
$i^\textrm{th}$ peak, and $C^B$ is a constant base-line offset.  We
attribute the deviation of the experimental line shape from the
expected Lorentzian line shape to residual misalignment of the two
beams, which we discuss further below.  We designate the resonant
repetition-rate frequency for the $i^\textrm{th}$ peak determined by fitting the calculated spectra for a given $\nu_{6\textrm{S}_{1/2}:
n^{\prime\prime} L^{\prime\prime}_{J^{\prime\prime}}}$,
$A_{n^{\prime\prime} L^{\prime\prime}_{J^{\prime\prime}}}$, and
$B_{n^{\prime\prime} L^{\prime\prime}_{J^{\prime\prime}}}$  by
$f_i^{calc}\prn{\nu_{6\textrm{S}_{1/2}: n^{\prime\prime}
L^{\prime\prime}_{J^{\prime\prime}}},A_{n^{\prime\prime}
L^{\prime\prime}_{J^{\prime\prime}}},B_{n^{\prime\prime}
L^{\prime\prime}_{J^{\prime\prime}}}}$.  The resonant
repetition-rate frequencies determined from the experimental spectra we designate as $f_i^{expt}$.  The best fit values for two-photon
frequency and the hyperfine constants were determined by minimizing
the weighted $\chi^2$ function
\begin{widetext}
\begin{align}
&\chi^2\prn{\nu_{6\textrm{S}_{1/2}: n^{\prime\prime}
L^{\prime\prime}_{J^{\prime\prime}}},A_{n^{\prime\prime}
L^{\prime\prime}_{J^{\prime\prime}}},B_{n^{\prime\prime}
L^{\prime\prime}_{J^{\prime\prime}}}} =
\sum_i\prn{\frac{f_i^{expt}-f_i^{calc}\prn{\nu_{6{\rm S}_{1/2}:
n^{\prime\prime}
L^{\prime\prime}_{J^{\prime\prime}}},A_{n^{\prime\prime}
L^{\prime\prime}_{J^{\prime\prime}}},B_{n^{\prime\prime}
L^{\prime\prime}_{J^{\prime\prime}}}}}{\sigma_i^{expt}}}^2 \label{eq
chisquared}
\end{align}
\end{widetext}
with respect to $\nu_{6{\rm S}_{1/2}: n^{\prime\prime}
L^{\prime\prime}_{J^{\prime\prime}}}$, $A_{n^{\prime\prime}
L^{\prime\prime}_{J^{\prime\prime}}}$, and $B_{n^{\prime\prime}
L^{\prime\prime}_{J^{\prime\prime}}}$.  Here $\sigma_i^{expt}$ is
the standard uncertainty in $f_i^{expt}$ determined by the fit to
the experimental data. For $N$ peaks and all three parameters
($\nu_{6{\rm S}_{1/2}: n^{\prime\prime}
L^{\prime\prime}_{J^{\prime\prime}}}$, $A_{n^{\prime\prime}
L^{\prime\prime}_{J^{\prime\prime}}}$, and $B_{n^{\prime\prime}
L^{\prime\prime}_{J^{\prime\prime}}}$) one would expect the minimum
value of the $\chi^2$ to be $N-3$.  We consistently observed larger
values of the $\chi^2$.  We attribute this uncertainty to residual
misalignment, which will be discussed in more detail below.  In
order to account for this larger $\chi^2$ value, we normalize the
$\chi^2$ function so that its minimum corresponds to the expected
value.  We then determine the uncertainty in the two-photon optical
frequency and the hyperfine coupling constants by finding the values
of $\nu_{6{\rm S}_{1/2}: n^{\prime\prime}
L^{\prime\prime}_{J^{\prime\prime}}}$, $A_{n^{\prime\prime}
L^{\prime\prime}_{J^{\prime\prime}}}$, and $B_{n^{\prime\prime}
L^{\prime\prime}_{J^{\prime\prime}}}$ where the normalized $\chi^2$
function increases by one from the point at which it is a minimum.
This amounts to increasing the uncertainty to make the frequencies
extracted from the different peaks self consistent.

\subsection{Results for the \eightS state}\label{sec 8Ssyst}

The spectral range of the comb allowed the \eightS state to be
studied by use of transitions through either the \sixPone (895 nm
and 761 nm) or \sixPthree (852 nm and 795 nm) states.  This
redundancy provided a check for consistency of the measurement and
the parameter extraction method. A complete study of the systematics was performed only for the transitions that occurred through the
\sixPthree states.

\begin{figure}
\centerline{\includegraphics[width=3.25 in]{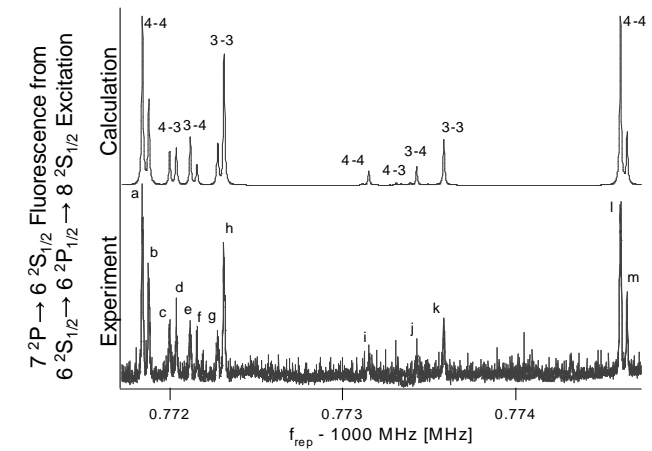}}
\caption{The calculated spectrum (upper trace) and experimental
signal (lower trace) of the $7\textrm{P} \rightarrow 6\textrm{S}$
fluorescence when the Cs atoms are excited to the \eightS state
through the \sixPone state.  The peaks in the experimental spectrum
that are labeled with letters were used in extracting the transition
frequencies as described in the text.} \label{fig 8SD1Spec}
\end{figure}

Figure \ref{fig 8SD1Spec} shows the spectrum for the
\sixS$\rightarrow$\sixPone$\rightarrow$\eightS transition along with the calculated spectrum.  The spectrum was collected by use of a 890 nm filter with a 10 nm passband for F1 and a combination of a 755 nm filter with a 40 nm pass band with a long-pass filter at 715 nm for F2.  The 13 peaks, labeled $a$ through
$m$ in Fig. \ref{fig 8SD1Spec}, were used to extract the transition
frequencies.  The peaks were acquired with both increasing and
decreasing scans of $f_{rep}$, resulting in 26 individual peaks. By
use of the procedure described above, the extracted
center-of-gravity two-photon frequency and the hyperfine $A$
constant for the \eightS state were found to be $729\, 009\,
798.844(38)$ MHz and $219.137(19)$ MHz. The $\chi^2$ was 31 for 26
peaks.  Here the uncertainties are determined from the $\chi^2$
function, as described in Section \ref{sec analysis}, and do not include any systematic uncertainties.

The \sixS$\rightarrow$\sixPthree$\rightarrow$\eightS transition is
shown in Fig.\ \ref{fig 8SD2Spec}. For this spectrum, one of the
counter-propagating laser beams was filtered with an interference
filter centered at 850~nm with a 10~nm pass band.  This resulted in
an average intensity of  $\approx 60$ $\textrm{W/cm}^2$ in the
interaction region, spread over $\approx 4000$ optical modes.  This
light served to excite the first step of the two-photon transition
at 852 nm.  Based on the transmission profile of the filter we
estimate that we have $\approx 10$ $\textrm{mW/cm}^2$ of light in
the optical mode resonant with the \sixS$\rightarrow$\sixPthree
transition.  The second stage of the transition was selected by use
of a combination of a 800 nm short-pass filter and a 780 nm
long-pass filter and provided $\approx$ 70 $\textrm{W/cm}^2$ over
$\approx 8000$ modes.  Using the procedure described above we fit
the 14 peaks labeled in the figure and find a center-of-gravity
frequency of $729\, 009\, 798.863(29)$ MHz and a hyperfine coupling
constan $219.133(17)$ MHz, where again, the uncertainties are
determined purely from the $\chi^2$ function. The value of the
$\chi^2$ was 64 for 28 peaks (14 peaks from both the increasing and
decreasing frequency scans) prior to the normalization. As described
in Section \ref{sec analysis} the uncertainties are increased to
make the $\chi^2$ consistent with the expected value. The resulting
frequencies are in good agreement with the value extracted from the
transition through the \sixPone state.

We analyzed in detail the systematic effects that could lead to a
shift of the measured frequencies from the true transition
frequencies using the transitions to the \eightS state through the
\sixPthree state.  Systematic shifts might be more significant for
excitation through the \sixPthree state, because of the smaller
hyperfine structure shifts of the \sixPthree state, relative to the
\sixPone state.  In addition, the signal-to-noise ratio of the
spectrum excited through the \sixPthree state was higher than that
excited through the \sixPone state, as an artifact of the filters
used in the measurements.

We considered four possible sources of systematic effects: ac Stark
shifts, Zeeman shifts, pressure shifts, and errors arising from
misalignment of the counter-propagating laser beams. The associated
uncertainties are summarized in Tab.\ \ref{tab ErrorBudget} and
described below.

To evaluate the systematics we focused on the peaks labeled $d$,
$e$, and $f$ in Fig.\ \ref{fig 8SD2Spec}.  A typical data set and
fit to the peaks is shown in Fig.\ \ref{fig 8SD2Sample}.  The effect of ac Stark shifts on the extracted frequencies was investigated by
varying the optical powers.  Neutral density filters were used to
reduce the power in both beams by approximately the same amount.
Data were collected at three optical powers ranging from the maximum
power, which was used for collection of the full spectrum shown in
Fig.\ \ref{fig 8SD2Spec}, to $60\:\%$ of the maximum power.  The
three peaks used in this analysis correspond to the \sixS$\prn{F =
4}\rightarrow$\sixPthree$\prn{F^\prime =
3,4,5}\rightarrow$\eightS$\prn{F^{\prime\prime}=4}$ transitions.
Because the final state is the same for all three peaks, it is not
possible to extract both the center-of-gravity frequency and the
hyperfine $A$ constant from these data.  In order to place a limit
on possible effects due to the ac Stark shifts on the
center-of-gravity frequency, we fixed the hyperfine $A$ constant to
the value determined from the analysis of the full spectrum and then
extracted a center-of-gravity frequency at each power by finding the
minimum in the $\chi^2$ function as defined by Eq.\ \eqref{eq
chisquared}. The center-of-gravity frequencies as a function of
power were fitted to a straight line.  We used the slope of the line
to look for any possible dependence of the center-of-gravity
frequency on the power. We find the slope to be $-43(123)$
$\textrm{kHz}/P$, where $P$ is the normalized operating power.
Combining the value of the slope with its uncertainty we arrive at a
maximum possible shift of 166 kHz at the the nominal operating
power, $P=1$.  We take this value as an estimate of the systematic
uncertainty from the ac Stark effect. Because we fix the hyperfine
$A$ constant this approach provides an upper limit on the
center-of-gravity frequency. Therefore, we do not attempt to apply a
correction to the data. To determine the possible effect on the
hyperfine splitting we fix the center-of-gravity frequency to the
value determined from the analysis of the full spectrum and extract
the hyperfine $A$ constant from the data. Fitting the data for the
$A$ constant we found a slope of $-22(72)$ $\textrm{kHz}/P$,
resulting in an uncertainty of 94 kHz.

\begin{figure}
\centerline{\includegraphics[width=3.5 in]{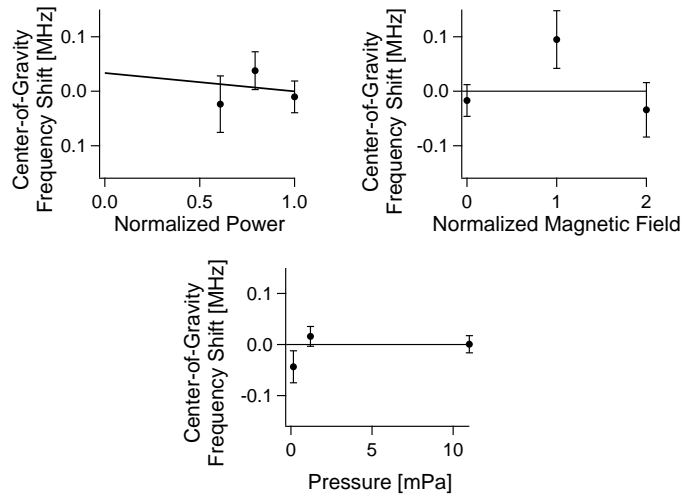}}
\caption{The dependence of the center-of-gravity frequencies for the
\sixS$\rightarrow$\sixPthree$\rightarrow$\eightS state on the power
(upper left), the magnetic field (upper right), and the Cs vapor
pressure (bottom). The straight line for the power dependence is a
weighted linear fit to the data.  The vertical axis has been offset
and centered so that the nominal operating point is at zero for the
power-dependence plot. The magnetic field and pressure data are
shown scattered about the weighted mean of the data. The horizontal
axis for the power dependence is plotted as a function of the
fraction of the maximum power. The horizontal axis for the magnetic
field dependence is plotted as a function of the residual magnetic
field, $\approx 50\: \mu T$.} \label{fig 8SD2Systematics}
\end{figure}

For two-photon transitions between S states with $\Delta M_F=0$
there is no linear Zeeman shift.  However, the laser polarization is
not perfectly linear and these states are step-wise resonant through
the intermediate P states, which are magnetically sensitive.  In
addition, the power on the first stage of the transition is near
saturation, so there may be some optical pumping that could couple
with an external magnetic field to lead to asymmetric line shapes.
To investigate the effect of Zeeman-shifts due to imperfect
cancelation of stray magnetic fields, data were collected without
the compensation coils and also with the current through the
compensation coils reversed, thus doubling the residual magnetic
field. Again, the peaks labeled $d$, $e$, and $f$ in Fig.\ \ref{fig
8SD2Spec} were used.  The value of the hyperfine $A$ constant was
fixed and the center-of-gravity frequency was extracted as described
above.  We see no evidence for a systematic shift as the magnetic
field is increased (Fig.\ \ref{fig 8SD2Systematics}).  The standard
deviation of the data is 70~kHz, which we take as the uncertainty
associated with the magnetic field.  A similar analysis for the
hyperfine $A$ constant (with the center-of-gravity frequency fixed)
also shows no evidence of any shift with the magnetic field. Using
the standard deviation of the points shown in Fig.\ \ref{fig
8SD2Systematics} as an estimate of the uncertainty we assign a
40~kHz uncertainty to the $A$ constant due to possible magnetic
fields.

For an analysis of the shifts due to collisions, the pressure of the Cs vapor was changed by heating the vapor cell as described above.
Data were taken at temperatures of 297, 318.6, and 345.5~K, corresponding to Cs vapor pressures of 0.15, 1.2, and 11~mPa.  As described above, we determine the
center-of-gravity frequency with a fixed hyperfine $A$ constant for
each pressure.  We see no indication of any variation of the data
with the Cs vapor pressure (Fig.\ \ref{fig 8SD2Systematics}). The
standard deviation of the frequencies extracted at the different
temperatures is 31 kHz, which we take as the uncertainty in the
center-of-gravity frequency due to temperature or pressure effects.
For the hyperfine $A$ constant, the scatter in the data taken at
different temperatures gives a standard deviation of 18 kHz, which
we adopt as the uncertainty.

The final systematic considered was possible error due to imperfect
alignment of the counter-propagating laser beams.  The effect of
misalignment of the two beams is to broaden and shift the peaks
through a first-order Doppler shift.  As can be seen from Eq.\
\eqref{eq twophotonprob}, the position of the peaks depends on the
difference in the magnitude and direction of the two vectors
$\vec{k}_1$ and $\vec{k}_2$ as well as the velocity class excited.
Additional first-order Doppler shifts come into play if the
wavefronts of $\vec{k}_1$ and $\vec{k}_2$ are not precisely
counter-propagating, which can result from divergence of the
counter-propagating beams and mismatch of the spatial modes. For a
spectrum that contains many peaks coming from many different
velocity classes the shift due to the misalignment largely averages
out. This effect was investigated theoretically by modifying the
program to include the effect of misalignment of the two beams. This
was done by changing the $\vec{k}_2\cdot \vec{v}$ term in Eq.\
\eqref{eq twophotonprob} to include an additional perpendicular
velocity component.  For the 8S state three spectra were calculated
with different angles of misalignment.  Analysis of the calculated
spectrum and an estimate of the maximum possible misalignment of the
beams gives a conservative limit on possible effects from angular
misalignment of 22 kHz for the center-of-gravity frequency and 20
kHz for the hyperfine $A$ constant. If we attribute the scatter in
the experimental results to angular misalignment, we conclude that
for all the spectra, the effective angular deviation due to either
misalignment or to divergence and mode mismatch of the beams is less
than 0.04 radians.  We note that for the calculated data the
misalignment of the counter-propagating laser beams leads to no
significant shift in the extracted frequency, it does lead to a
larger spread of the data and an increased value of  $\chi^2$, as
observed. This additional spread of the data beyond what is expected
from the statistical noise is taken into account by conservatively
increasing the uncertainties to achieve the expected $\chi^2$, as
described in Section \ref{sec analysis}.

In addition, the effect of misalignment was investigated
experimentally on the
\sixS$\rightarrow$\sixPone$\rightarrow$\sevenDthree transition. The
study of this transition confirmed that while the individual peaks
do shift, there was no shift in the mean value of the transition
frequency, within the uncertainty of our measurement.

The non-zero second-order Doppler shift $(v/c)^2\, \nu_0$ is about
300 Hz for this transition and is negligible compared to the other
uncertainties.

Table \ref{tab ErrorBudget} summarizes the corrections and
associated uncertainties on the extracted center-of-gravity and
hyperfine $A$ constants.  We see no evidence for any shift that is
larger than the uncertainty associated with evaluating the effect.
As our final value we take the mean of the \eightS frequencies
determined from the \sixPone and \sixPthree intermediate states,
weighted by their respective statistical uncertainties, and apply
the corrections and systematic uncertainties (propagated in
quadrature) listed in Table \ref{tab ErrorBudget}. We find final
values for the center-of-gravity frequency and hyperfine $A$
constant of \eightS state of
\begin{align}
\nu_{\textrm{6S}_{1/2}:\textrm{8S}_{1/2}} & = 729
\, 009\, 798.86(19) \: \textrm{MHz} \nonumber \\
A_{\textrm{8S}_{1/2}}& =219.14(11)\:  \textrm{MHz} \nonumber .
\end{align}
These results are in good agreement with previous measurements of
the \eightS state, summarized in Table \ref{tab
previousMeasurements}.


\subsection{Results for the \nineS state}

\begin{figure}
\centerline{\includegraphics[width=3.25 in]{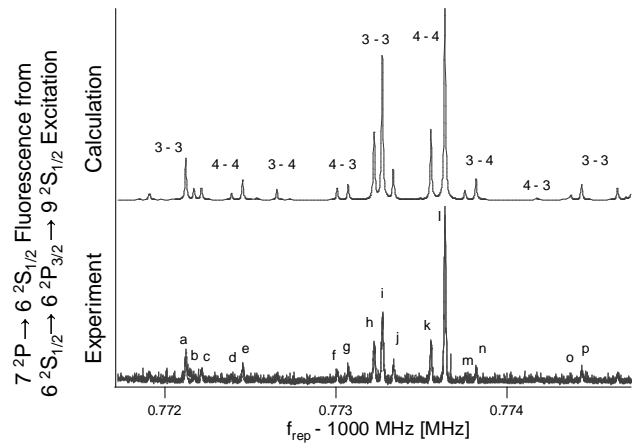}}
\caption{The calculated spectrum (upper trace) and experimental
signal (lower trace) of the $7\textrm{P} \rightarrow 6\textrm{S}$
fluorescence when the Cs atoms are excited to the \nineS state
through the \sixPthree state.  The peaks in the experimental
spectrum labeled with letters were used in extracting the transition
frequencies, as described in the text.} \label{fig 9SD2Spec}
\end{figure}

Figure \ref{fig 9SD2Spec} shows the spectrum obtained for the
cascaded two-photon transition
\sixS$\rightarrow$\sixPthree$\rightarrow$\nineS and the calculated
spectrum.  This spectrum was collected by use of a 850 nm filter with a 10 nm passband for F1 and a 658 nm filter with a 10 nm passband for F2. The measurement of the \nineS state through the \sixPone intermediate state was not possible because the complementary radiation at 635.5 nm was on the edge of the laser bandwidth.


The peaks $a$-$p$ were used to determine the center-of-gravity
frequency and the hyperfine $A$ constant as described above. The
extracted center-of-gravity frequency is $806\, 761\, 363.380(58)$
MHz and the hyperfine $A$ constant is $109.932(30)$ MHz.  The
$\chi^2$ was 83 for 32 peaks.

The systematic effects due to ac-Stark shifts, Zeeman shifts,
pressure shifts, and misalignment effects were investigated in the
same way as they were for the \eightS state. The systematic effects
were studied by focusing on the three peaks labeled $h$, $i$, and
$j$ in Fig.\ \ref{fig 9SD2Spec}, corresponding to transitions from
the $F = 3$ ground state hyperfine level to the $F^{\prime\prime} =
3$ excited state hyperfine level through the $F^\prime = 2$, 3, and
4 intermediate hyperfine levels.

\begin{figure}
\centerline{\includegraphics[width=3.25 in]{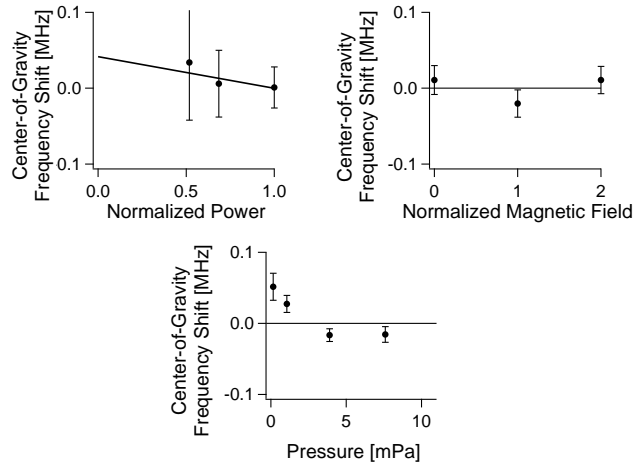}}
\caption{The dependence of the center-of-gravity frequencies for the
\sixS$\rightarrow$\sixPthree$\rightarrow$\nineS state on the power
(upper left), the magnetic field (upper right), and the Cs vapor
pressure (bottom). The straight line for the power dependence is a
weighted linear fit to the data.  The vertical axis has been offset
and centered so that the nominal operating point is at zero for the
power-dependence plot. The magnetic field and pressure data are
shown scattered about the weighted mean of the data. The power is
plotted as a function of the fraction of the maximum power. The
magnetic field is plotted as a function of the residual magnetic
field, $\approx 50\: \mu T$.} \label{fig 9SD2Systematics}
\end{figure}

The laser intensity for both transitions was varied.  For the
\sixS$\rightarrow$\sixPthree stage of the transition the variation
was from $\approx 40\: \textrm{W/cm}^2$ to $20 \: \textrm{W/cm}^2$
and was spread over $\approx 4000$ optical modes. A linear fit of
the center-of-gravity frequency to the power gave a dependence of
$-42(127)$ $\textrm{kHz}/P$ where $P$ is the normalized operating
power.  We therefore assign a 169 kHz uncertainty to the
center-of-gravity frequency due to possible light shifts (Fig.\
\ref{fig 9SD2Systematics}).  Similarly, for the hyperfine $A$
constant we found a linear dependence consistent with zero slope,
$21(58)$ $\textrm{kHz}/P$.  We take 79 kHz uncertainty to be the
uncertainty in the $A$ constant due to possible ac-Stark shifts.

As with the \eightS state, no discernable dependence of the
frequency on the external magnetic field was observed (Fig.\
\ref{fig 9SD2Systematics}).  Based on the scatter (standard
deviation) of the data we arrive at an uncertainty of 18 kHz for the center-of-gravity frequency and 7 kHz for the hyperfine $A$
constant.

Spectra were taken at four different temperatures of the vapor cell corresponding to Cs vapor pressures of 0.15, 1.06, 3.9, and 7.6~mPa.  The dependence of the  center-of-gravity frequency is shown in Fig.\ \ref{fig 9SD2Systematics}.  There is evidence of a possible shift in the frequency with pressure.  However, even at pressures that are 50 times larger than the nominal operating pressure, we see only a
67~kHz shift.  We again use the standard deviation of the data to
place a conservative limit on possible uncertainties due to pressure shifts and find a 34 kHz uncertainty in the optical frequency due to temperature dependent effects.  A similar analysis for the hyperfine $A$ constant gives an uncertainty due to temperature dependent
effects of 15 kHz.

An analysis of the frequency dependence on misalignment done in the
same way as described above yielded a 58 kHz uncertainty for the
center-of-gravity frequency and 20 kHz for the hyperfine $A$
constant.

The uncertainties to the center-of-gravity frequency and the
hyperfine $A$ constant are summarized in Table \ref{tab
ErrorBudget}.

We arrive at a final value for the center-of-gravity frequency for
the \sixS$\rightarrow$\nineS transition and the hyperfine $A$
constant of the \nineS state of
\begin{align}
\nu_{\textrm{6S}_{1/2}:\textrm{9S}_{1/2}} & = 806
\, 761\, 363.38(19) \: \textrm{MHz} \nonumber \\
A_{\textrm{9S}_{1/2}}& =109.93(9)\:  \textrm{MHz} \nonumber .
\end{align}
This is in good agreement with the previous measurements of this
transition, summarized with other data in Table \ref{tab
previousMeasurements}, and improves on the center-of-gravity
frequency by two orders of magnitude.

\subsection{Results for the \sevenDthree state}\label{sec
sevenDthree}

\begin{figure}
\centerline{\includegraphics[width=3.25 in]{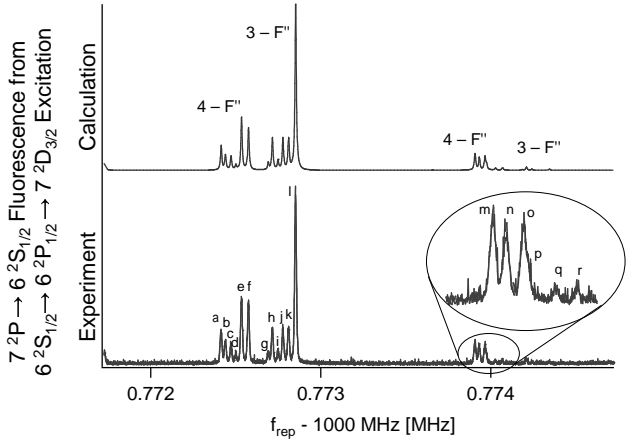}}
\caption{The calculated spectrum (upper trace) and experimental
signal (lower trace) of the $7\textrm{P} \rightarrow 6\textrm{S}$
fluorescence when the Cs atoms are excited to the \sevenDthree state
through the \sixPone state.  The peaks in the experimental spectrum
labeled with letters were used in extracting the transition
frequencies as described in the text.} \label{fig 7DD1Spec}
\end{figure}

The center-of-gravity frequency, magnetic dipole, and electric
quadrupole hyperfine coupling constants for the \sevenDthree state
were determined from the spectrum in Fig.\ \ref{fig 7DD1Spec}.  The filters used in collecting this spectrum were a 890 nm filter with a 10 nm passband for F1 and a 670 nm filter with a 30 nm passband for F2. This spectrum was taken by exciting Cs through the \sixPone intermediate state. Use of the \sixPone state eliminates possible transitions to the \sevenDfive state due to angular momentum selection rules.
The groups consist of six peaks, corresponding to the $F\rightarrow F^\prime=3\rightarrow F^{\prime\prime}=2,3,4$ and $F\rightarrow F^\prime=4\rightarrow F^{\prime\prime}=3,4,5$ transitions.  The 18 peaks labeled $a$ through $r$ in Fig.\ \ref{fig
7DD1Spec} were used to extract the frequency and hyperfine coupling
constants.

The 7D state has $J>1/2$ and, consequently, an electric quadrupole
coupling denoted by the hyperfine $B$ constant.  Minimizing the
$\chi^2\prn{\nu,\, A,\, B}$ function defined by Eq.\ \eqref{eq
chisquared} with respect to the center-of-gravity frequency and the
hyperfine $A$ and $B$ constants, we find a value for the center of
gravity of $780\, 894\, 762.250(28)$ MHz, a hyperfine $A$ constant
value of $7.386(6)$ MHz, and a hyperfine $B$ constant value of
$-0.182(50)$.  The $\chi^2$ was $78$ for 36 peaks.

\begin{figure}
\centerline{\includegraphics[width=3.25 in]{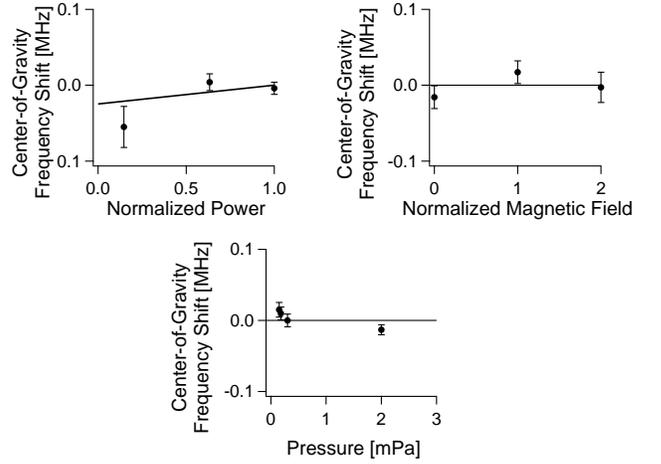}}
\caption{The dependence of the center-of-gravity frequencies for the
\sixS$\rightarrow$\sixPone$\rightarrow$\sevenDthree transition on
the power (upper left), the magnetic field (upper right), and the Cs
vapor pressure (bottom).  The straight line for the power dependence
is a weighted linear fit to the data.  The vertical axis has been
offset and centered so that the nominal operating point is at zero
for the power-dependence plot. The magnetic field and pressure data
are shown scattered about the weighted mean of the data. The power
is plotted as a function of the fraction of the maximum power. The
magnetic field is plotted as a function of the residual magnetic
field, $\approx 50\: \mu T$.} \label{fig 7DD1Systematics}
\end{figure}

The systematic effects were studied by focusing on the six peaks
labeled $g$, $h$, $i$, $j$, $k$, and $l$ in Fig.\ \ref{fig
7DD1Spec}, corresponding to transitions from the $F=3$ ground state.  The laser intensity on both stages of the transition was
varied. For the \sixS$\rightarrow$\sixPone stage, the change in
intensity ranged from $\approx 19\: \textrm{W/cm}^2$ to $\approx 3\:
\textrm{W/cm}^2$, spread over $\approx 4000$ optical modes within
the 10 nm (3.8 THz) filter bandwidth.  The slope of the
center-of-gravity frequency as a function of light power yields a
dependence of $22(27)$ kHz (Fig.\ \ref{fig 7DD1Systematics}),
corresponding to an uncertainty of 49 kHz.  We find uncertainties
due to ac-Stark shifts of 14 kHz for the $A$ constant and $17$ kHz
for the $B$ constant.

No effect of the external magnetic field was observed (Fig.\
\ref{fig 7DD1Systematics}), within the measurement uncertainties.
Based on the scatter of the data, we assign uncertainties of 17 kHz,
3 kHz, and 17 kHz for the center-of-gravity frequency, the hyperfine
$A$ constant and the hyperfine $B$ constant, respectively.

Spectra were taken at four different temperatures of the vapor cell, corresponding to a Cs vapor pressure of 0.15, 0.2, 0.3, and 2.0~mPa.  As with the \nineS state, there appears to be a small dependence of the center-of-gravity frequency as a function of the pressure in the cell, leading to a 28 kHz shift when the pressure is increased by a
factor of $\approx 10$.  However, since we fix the $A$ and $B$
constants, this serves as a maximal shift in the center-of-gravity
frequency and we do not attempt to correct for this shift.  We take
the standard deviation as a conservative estimate of the error and
assign an uncertainty of 12~kHz to the center-of-gravity frequency
due to possible temperature-dependent effects.  The hyperfine $A$
constant shows no such dependence and the scatter is less than the
uncertainty on the mean. We therefore use the uncertainty on the
weighted mean of the data taken at different temperatures to arrive
at an uncertainty of 1 kHz for the hyperfine $A$ constant. For the
hyperfine $B$ constant we find an 11 kHz uncertainty (from the
scatter of the data).

A theoretical analysis of the frequency dependence on misalignment
similar to that described for the \eightS state analysis yielded a 48
kHz uncertainty for the center-of-gravity frequency and
uncertainties of 12~kHz for the hyperfine $A$ constant and 143 kHz
for the $B$ constant. In addition, the effects of misalignment were
investigated experimentally.  Spectra of the six peaks labeled $g$,
$h$, $i$, $j$, $k$, and $l$ in Fig.\ \ref{fig 7DD1Spec} were taken
at three different beam alignments. The degree of misalignment was
characterized by the peak amplitude. Misalignment of the beams so
that the peak amplitudes decreased by more than a factor of two
resulted in no detectable change in the center-of-gravity, $A$
constant, or $B$ constant.

The uncertainties to the center-of-gravity frequency and the
hyperfine $A$ constant are summarized in Table \ref{tab
ErrorBudget}.


With the uncertainties from Table \ref{tab ErrorBudget} we find
\begin{align}
\nu_{\textrm{6S}_{1/2}:\textrm{7D}_{3/2}} & = 780\, 894\, 762.250(77) \: \textrm{MHz} \nonumber \\
A_{\textrm{7D}_{3/2}}& =7.386(15)\:  \textrm{MHz} \nonumber \\
B_{\textrm{7D}_{3/2}}& =-0.18(16)\:  \textrm{MHz} \nonumber.
\end{align}
These numbers are in good agreement with previous measurements,
shown in Table \ref{tab previousMeasurements} and lead to a
significant improvement in the knowledge of the center-of-gravity
frequency.

\subsection{Results for the \sevenDfive state}
\begin{figure}
\centerline{\includegraphics[width=3.25 in]{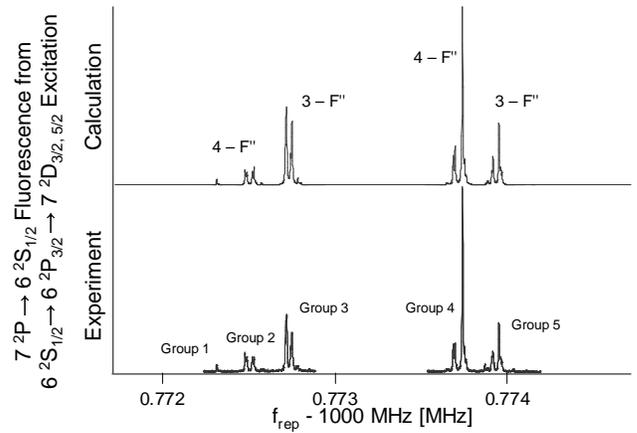}}
\caption{The calculated spectrum (upper trace) of the $7\textrm{P}
\rightarrow 6\textrm{S}$ fluorescence when the Cs atoms are excited
to the ${\rm 7D}_{3/2, \, 5/2}$ state.  The experimental signal
(lower trace) shows the $7\textrm{P} \rightarrow 6\textrm{S}$
fluorescence when the Cs atoms are excited to the both the
$7\textrm{D}_{3/2, \, 5/2}$ states through the \sixPthree state. In
the experimental spectrum data are shown for only those regions
where there was a signal.} \label{fig 7DD2Spec}
\end{figure}

\begin{figure}
\centerline{\includegraphics[width=3.25 in]{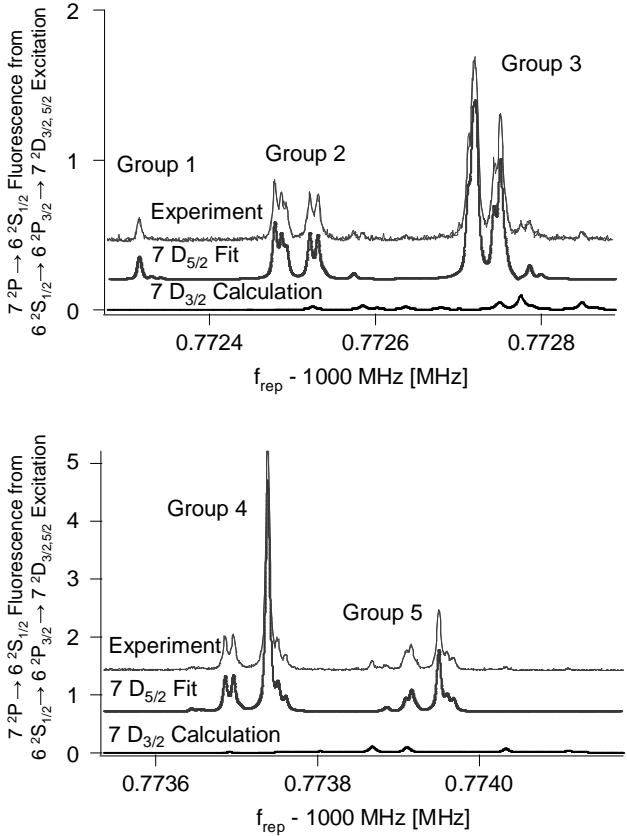}}
\caption{The experimental spectrum (upper trace), fit to the
\sixS$\rightarrow$\sixPthree$\rightarrow$\sevenDfive transition
(middle trace), and of the calculated contribution from the
\sixS$\rightarrow$\sixPthree$\rightarrow$\sevenDthree transition
(lower trace).  Note that the vertical scales are different for the
two figures.} \label{fig 7D32And7D52}
\end{figure}

Figure \ref{fig 7DD2Spec} shows the experimental fluorescence
spectra for excitation of the \sevenDthree and \sevenDfive states
through the \sixPthree state along with the calculated spectra for
excitation to the \sevenDthree and \sevenDfive states.  The spectrum was collected using a 850 nm filter with a 10 nm passband for F1 and a 700 nm filter with a 25 nm passband for F2.  The close
energy spacing of the \sevenDthree and \sevenDfive states did not
allow for a selection of only one of the states when excited through
the \sixPthree state. However, because we were able to determine the
\sevenDthree state energy and hyperfine splitting as described in
Section \ref{sec sevenDthree}, we were able to fix the frequency of
the \sevenDthree state from these measurements and use the spectrum
shown in Fig.\ \ref{fig 7DD2Spec} to extract information about the
\sevenDfive state.  In addition, angular momentum considerations
result in much stronger transition probabilities for the
\sixPthree$\rightarrow$\sevenDfive transition compared to the
\sixPthree$\rightarrow$\sevenDthree transition. To remove the effect
of the overlap, we calculated the \sevenDthree spectrum through the
\sixPthree state using the values determined by the
\sixS$\rightarrow$\sixPone$\rightarrow$\sevenDthree excitation.
Figure \ref{fig 7D32And7D52} shows the contribution of the
transitions to the \sevenDthree to the fluorescence spectrum.  We
superposed that calculated spectrum on the experimental spectrum and
adjusted the peak intensities to match the isolated \sevenDthree
peaks.  In this way we accounted for and removed the \sevenDthree
parameters. To quantify the influence of the \sevenDthree state and
estimate the uncertainty resulting from the overlapped spectra, we
fit the spectra with and without subtracting the spectrum from the
\sevenDthree state. We found a maximum deviation of 10 kHz in the
center-of-gravity frequency of the \sevenDfive state, 1 kHz in the
hyperfine $A$ constant, and 30 kHz in the hyperfine $B$ constant
which we take to be the uncertainties due to this overlap. The
contributions from the two transitions are separated in Fig.\
\ref{fig 7D32And7D52}.

For a given ground state hyperfine component $F$, there are nine
distinct transitions due to the manifold of intermediate states and
upper states.  The five groups of peaks labeled in Fig.\ \ref{fig
7DD2Spec} were used to extract values for the center-of-gravity and
hyperfine $A$ and $B$ constants.  In group one, only three peaks
were of sufficient amplitude to be fit. All nine peaks were fit in
groups three and five.  Seven peaks were used in group two and six
peaks in group four.  From the fit to these 34 peaks we determined
the center-of-gravity frequency to be $781\, 522\, 153.682(25)$ MHz
and the hyperfine $A$ and $B$ constants were $-1.717(4)$ MHz and
$-0.182(86)$ MHz, respectively. To improve the precision of the
fits, we use scans with increasing frequency and decreasing
frequency simultaneously. The minimum $\chi^2$ was $90$ for 34 peaks
prior to normalization.

The systematics were studied in the same way as with the previous
states.  We focused on group four of Fig.\ \ref{fig 7D32And7D52} due
to its size and also the fact that there is very little contribution
from the \sevenDthree state for this group. The optical intensity
was varied for both stages of the transition.  For the first stage
the intensity over the 10 nm bandwidth was varied from $\approx 22\:
\textrm{W/cm}^2$ to $\approx 5\: \textrm{W/cm}^2$. Data were taken
at cell temperatures corresponding to saturated vapor pressures of
0.15, 1.05, 3.9, and 8.5 mPa.  The dependence of the
center-of-gravity on the varied parameters is shown in Fig.\
\ref{fig 7DD2Systematics} and are summarized in Table \ref{tab
ErrorBudget}.

As with the previous transitions there was no significant shift
arising from the light power.  A linear fit gave a slope of
$-23(62)$ $\textrm{kHz}/P$, where $P$ is the normalized operating
power.  This gives us an uncertainty of 85~kHz in the
center-of-gravity frequency.  We found uncertainties in the
hyperfine $A$ and $B$ constants of 11~kHz and 458~kHz, respectively.

Unlike the other transitions studied here, we did observe a
dependence of the center-of-gravity on the magnetic field (Fig.\
\ref{fig 7DD2Systematics}).  There was clear evidence of broadening
in the peaks at the largest magnetic fields.  We attribute the shift
to a break down of the line-shape model as the line broadens coupled
with the overlap of the peaks.  In this case, we base our estimate
of the uncertainty on the difference of the point taken at the
zeroed field with the residual field and find an uncertainty of
35~kHz in the optical frequency.  There is no evidence for a
dependence of the extracted hyperfine $A$ constant, and we find a
magnetic-field uncertainty of 6 kHz for it.  For the $B$ constant we
find an uncertainty of 125 kHz.

We see no clear evidence for any temperature dependent effects.
Using the larger of standard deviation or the uncertainty in the
mean, we find uncertainties of 7~kHz, 1~kHz, and 38~kHz for the
temperature-dependent uncertainties of the center-of-gravity
frequency, the hyperfine $A$ constant, and the hyperfine $B$
constant, respectively

Applying these uncertainties we find
\begin{align}
\nu_{\textrm{6S}_{1/2}:\textrm{7D}_{5/2}} & = 781\, 522\, 153.68(16) \: \textrm{MHz} \nonumber \\
A_{\textrm{7D}_{5/2}}& =-1.717(15)\:  \textrm{MHz} \nonumber \\
B_{\textrm{7D}_{5/2}}& =-0.18(52)\:  \textrm{MHz} \nonumber.
\end{align}

\begin{figure}
\centerline{\includegraphics[width=3.25 in]{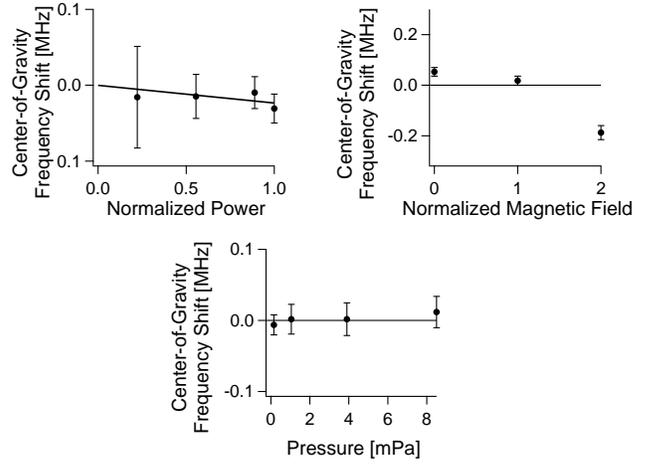}}
\caption{The dependence of the center-of-gravity frequencies for the \sixS$\rightarrow$\sixPthree$\rightarrow$\sevenDfive transition on
the power (upper left), the magnetic field (upper right), and the Cs vapor pressure (bottom). The straight line in the power dependence plot is a linear weighted fit to the data. The vertical axis has been offset and centered so that the nominal operating point is at zero. The power is plotted as a function of the fraction of the maximum power. The magnetic field is plotted as a function of the residual magnetic field.} \label{fig
7DD2Systematics}
\end{figure}


\section{Conclusions}\label{sec conclusions}

We have used the technique of velocity-selective two-photon
excitation to measure the center-of-gravity frequencies and the
hyperfine coupling constants for the \eightS, \nineS, \sevenDthree,
and \sevenDfive states of Cs.  The results for all of the
transitions are summarized and compared to other published
experimental values in Table \ref{tab previousMeasurements}.  This
technique has potential metrological applications for secondary
frequency standards at optical wavelengths.  At a more fundamental
level, the hyperfine coupling measurements are valuable in improving
atomic structure calculations.  The measurement accuracy of multiple
transitions at significantly different wavelengths rivals that of cw
laser spectroscopy techniques and demonstrates the versatility of
direct frequency comb spectroscopy.

Our measured transition frequencies are in good agreement with
previous measurements, while achieving greater accuracy for the
\nineS, \sevenDthree, and \sevenDfive states.  In these measurements
the uncertainties were typically limited by signal-to-noise ratio
and fitting uncertainties combined with uncertainties in the
determination of the ac Stark shifts and geometrical factors.

\begin{widetext}
\begin{center}
\begin{table}[h]
\caption{Summary of the contributions to the error budget for the
measured transition frequencies and hyperfine constants. All frequencies are given in kilohertz.} \label{tab
ErrorBudget}
\begin{tabular}{c|cc|cc|ccc|ccc}
\hline\hline
Effect & $\nu_{\textrm{6S}_{1/2}:\textrm{8S}_{1/2}}$ &$A_{\textrm{8S}_{1/2}}$ & $\nu_{\textrm{6S}_{1/2}:\textrm{9S}_{1/2}}$ &$A_{\textrm{9S}_{1/2}}$ & $\nu_{\textrm{6S}_{1/2}:\textrm{7D}_{3/2}}$ &
$A_{\textrm{7D}_{3/2}}$ & $B_{\textrm{7D}_{3/2}}$ &$\nu_{\textrm{6S}_{1/2}:\textrm{7D}_{5/2}}$ &
$A_{\textrm{7D}_{5/2}}$ & $B_{\textrm{7D}_{5/2}}$\\
\hline
ac Stark & 166 & 94 & 169 & 79 & 49 & 14 & 17 & 85 & 11 & 458\\
B-Field  & 70 & 40 &  18 &  7 & 17  & 3 & 17 & 35 & 6 & 125\\
Cs Vapor Pressure & 31  & 18 & 34 & 15 & 12 & 1 & 26 & 7  & 1 & 38\\
Alignment & 22 &  20 &  58 & 20 & 48 & 12 & 143 & 34 & 7 & 180 \\
Overlap of \sevenDthree & - & - & - & - & - & - & - & 9 & 1 & 30 \\
Statistical &  24 &  13 & 58 &  30 & 28 & 20 & 72 & 25 & 4 & 86\\
\hline Total & 186 &  106 & 192 & 88 & 77 &  15 & 164 & 102 &  15 & 517 \\\hline \hline
\end{tabular}
\end{table}
\end{center}

\begin{center}
\begin{table}[h]
\caption{Comparison of the center-of-gravity frequencies and
hyperfine coupling constants measured in this work with previous
measurements.  All numbers are in megahertz. The uncertainties
represent the standard error (68\% confidence interval).} \label{tab
previousMeasurements}
\begin{tabular}{lccc}
\hline\hline
Measurement & Center-of-Gravity Frequency & \hspace{0.05in}Hyperfine Constant $A$ \hspace{0.05in} & Hyperfine Constant $B$ \\
\hline \hline \eightS \\ \hline \hspace{0.125in} This Work & $729 \,
009\, 798.86(19)$ & 219.14(11) & - \\ \hspace{0.125in} Fendel
\textit{et al.} \cite{fendel07} & $729\, 009\, 799.020(26)$ &
219.125(4) & -
\\ \hspace{0.125in} Hagel \textit{et al.} \cite{hagel99} & $729\, 009\,
798.82(20)$ & 219.12(1) & -\\ \hline\hline \nineS \\ \hline
\hspace{0.125in}
This Work & $806\, 761\, 363.38(19)$ & 109.93(9) & - \\
\hspace{0.125in}
Weber \& Sansonetti \cite{weber87} & $806\, 761\, 372(8)$ & - & - \\
\hspace{0.125in}
Farley \textit{et al.} \cite{farley77} & - & 110.1(5) & - \\
\hspace{0.125in} Gupta \textit{et al.} \cite{gupta73} & - &
109.5(2.0) & - \\ \hline\hline \sevenDthree \\ \hline
\hspace{0.125in} This Work & $780 \, 894\, 762.250(77)$ & 7.386(15)
& -0.18(16) \\ \hspace{0.125in} Weber \& Sansonetti \cite{weber87}
& $780 \, 894\, 772(8)$ & - & - \\ \hspace{0.125in} Kortyna \textit{et al.} \cite{kortyna08} & - & 7.36(3) & -0.1(2) \\
\hspace{0.125in} Auzinsh \textit{et al.} \cite{auzinsh07} & - &
7.4(2) & - \\
\hline\hline \sevenDfive \\ \hline \hspace{0.125in} This Work &
$781\, 522\, 153.68(10)$ & -1.717(15) & -0.18(52)
\\ \hspace{0.125in} Weber \& Sansonetti \cite{weber87}
& $781 \, 522\, 153(8)$ & - & - \\ \hspace{0.125in} Auzinsh \textit{et al.} \cite{auzinsh07} & - &
-1.56(9) & - \\
\hline\hline
\end{tabular}
\end{table}
\end{center}
\end{widetext}

\section{Acknowledgements}

We thank Hugh Robinson for insightful advice and preparation of the
Cs vapor cell.  We are grateful to Jose Almaguer, Nate Newbury, and
Todd Johnson for carefully reading the manuscript and providing
valuable suggestions and discussions. One of us (VM) was supported
in part by the DST(RSA) PDP through the National Metrology Institute of South Africa (NMISA). CET acknowledges partial support from the
National Science Foundation through grant number PHY99-87984.

%

\bibliography{FrequencyComb}

\end{document}